\documentclass[10pt]{article}
\usepackage{amsmath}
\usepackage{booktabs}
\usepackage{fancyhdr}
\usepackage[dvipsnames]{xcolor}
\usepackage[most]{tcolorbox}

\usepackage[round]{natbib}
\usepackage{geometry}
\usepackage{graphicx}

\usepackage{charter}
\usepackage{hyperref}
\hypersetup{colorlinks=true,allcolors=BlueViolet}

\title{\textsc{\textbf{On-Demand Earth System Data Cubes}}}

\author{
David Montero\textsuperscript{1,2,*}\quad 
César Aybar\textsuperscript{3}\quad 
Chaonan Ji\textsuperscript{1}\quad
Guido Kraemer\textsuperscript{1}\\
Maximilian Söchting\textsuperscript{1}\quad
Khalil Teber\textsuperscript{1}\quad
Miguel D. Mahecha\textsuperscript{1,2,4}\\
\small\textsuperscript{1}RSC4Earth, IEF, Leipzig University\quad
\textsuperscript{2}iDiv\quad \textsuperscript{3}IPL, Universitat de Val\`encia \textsuperscript{4}UFZ\\
\small*Corresponding author: \texttt{david.montero@uni-leipzig.de}
}
\date{}

\begin{document}

\maketitle

\begin{abstract}
Advancements in Earth system science have seen a surge in diverse datasets. Earth System Data Cubes (ESDCs) have been introduced to efficiently handle this influx of high-dimensional data. ESDCs offer a structured, intuitive framework for data analysis, organising information within spatio-temporal grids.
The structured nature of ESDCs unlocks significant opportunities for Artificial Intelligence (AI) applications. By providing well-organised data, ESDCs are ideally suited for a wide range of sophisticated AI-driven tasks.
An automated framework for creating AI-focused ESDCs with minimal user input could significantly accelerate the generation of task-specific training data.
Here we introduce {\tt cubo}, an open-source Python tool designed for easy generation of AI-focused ESDCs.
Utilising collections in SpatioTemporal Asset Catalogs (STAC) that are stored as Cloud Optimised GeoTIFFs (COGs), {\tt cubo} efficiently creates ESDCs, requiring only central coordinates, spatial resolution, edge size, and time range.
\end{abstract}

\thispagestyle{fancy}
\renewcommand{\headrulewidth}{0pt}
\renewcommand{\footrulewidth}{0pt}
\fancyhf{}
\cfoot{\scriptsize{Copyright 2024 IEEE. Published in the 2024 IEEE International Geoscience and Remote Sensing Symposium (IGARSS 2024), scheduled for 7 - 12 July, 2024 in Athens, Greece. Personal use of this material is permitted. However, permission to reprint/republish this material for advertising or promotional purposes or for creating new collective works for resale or redistribution to servers or lists, or to reuse any copyrighted component of this work in other works, must be obtained from the IEEE. Contact: Manager, Copyrights and Permissions / IEEE Service Center / 445 Hoes Lane / P.O. Box 1331 / Piscataway, NJ 08855-1331, USA. Telephone: + Intl. 908-562-3966.}}

\section{Introduction}
\label{sec:intro}

Earth System Data Cubes (ESDCs) are multidimensional arrays encapsulating analysis-ready Earth system data, defined by their dimensions, grids, data, and attributes \citep{mahecha2020esdcs}. Recent advances in cloud technologies, such as the SpatioTemporal Asset Catalogs (STAC) specification, which simplifies geospatial data description and indexing; and Cloud Optimised GeoTIFF (COG), which allows for HTTP range requests; have enabled efficient generation of ESDCs from cloud-stored data \citep{montero2023avenues}. Generated ESDCs typically feature two spatial dimensions (such as $x$ and $y$), one temporal dimension, and the variable dimension. As ESDCs are usually cuboids, the length of the spatial grids can vary (e.g. global ESDCs have shorter latitude grids than longitude grids). In the case of Artificial Intelligence (AI) for local-scale applications, spatial grids of equal length are preferred for vision AI tasks. Examples include BigEarthNet's $120\times120$, $60\times60$, and $20\times20$ image patches \citep{sumbul2019bigearthnet}, and CloudSEN12's $509\times509$ image patches \citep{aybar2022cloudsen12}. We refer to ESDCs with spatial grids of equal length as ``AI-focused ESDCs''. Despite the availability of tools leveraging cloud technologies for ESDC creation, a systematic approach for producing AI-focused ESDCs on demand is lacking.

This paper introduces {\tt cubo}\footnote{https://github.com/ESDS-Leipzig/cubo}, an open-source Python-based tool streamlined for creating AI-focused ESDCs. {\tt cubo} enables automatic and minimal-input ESDC generation from cloud-stored data, greatly expanding the potential for generating comprehensive Earth system datasets. The paper is structured as follows: Sec.~\ref{sec:framework} details the {\tt cubo} framework and its simplification of AI-focused ESDC generation; Sec.~\ref{sec:showcase} presents examples of AI-focused ESDCs generated using {\tt cubo}; and Sec.~\ref{sec:conclusions} provides our conclusions.

\section{Framework}
\label{sec:framework}

To streamline {\tt cubo}'s functionality, we introduced a new ESDC characterisation (Sec.~\ref{sec:model}). This significantly simplifies the input process, requiring only a few input parameters from the user. Subsequently, {\tt cubo} utilises these user-defined parameters to construct an ESDC in a systematic manner (Sec.~\ref{sec:construction}).

\subsection{ESDC characterisation}
\label{sec:model}

\begin{figure}[!h]

    \centering
    \includegraphics[width=0.5\textwidth]{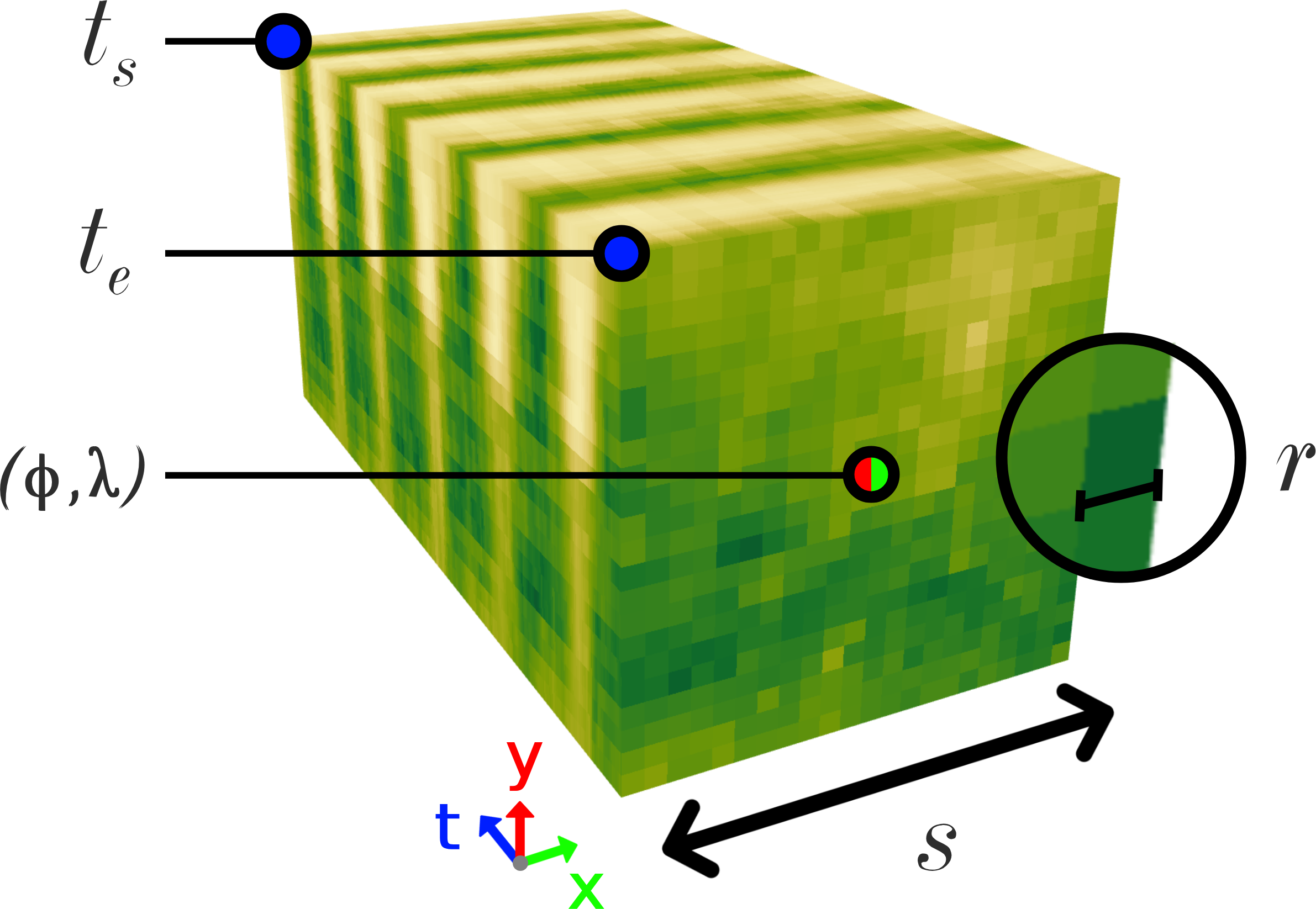}
    \caption{\textbf{{\tt cubo}'s ESDC characterisation}. Note that in this representation, $t_s$ and $t_e$ represent values from the temporal dimension, while $\phi$ and $\lambda$ represent coordinates associated with the spatial centre of the cube.}
    \label{fig:parameters}

\end{figure}

{\tt cubo} characterises AI-focused ESDCs using the parameters described in Box~1 and illustrated in Fig.~\ref{fig:parameters}.

\begin{tcolorbox}[colback=black!5,colframe=lightgray,title=Box 1: Parameters characterising ESDCs,fonttitle=\bfseries,coltitle=black]
\begin{itemize}
    \item Central coordinates of the cube, defined by latitude ($\phi$) and longitude ($\lambda$).
    \item Edge size of the cube in pixels ($s$).
    \item Spatial resolution in meters ($r$).
    \item Time range of the cube, defined by start ($t_s$) and end ($t_e$) timestamps.
\end{itemize}
\end{tcolorbox}

\subsection{ESDC construction}
\label{sec:construction}

\begin{figure*}[!h]
    \centering
    \includegraphics[width=1\textwidth]{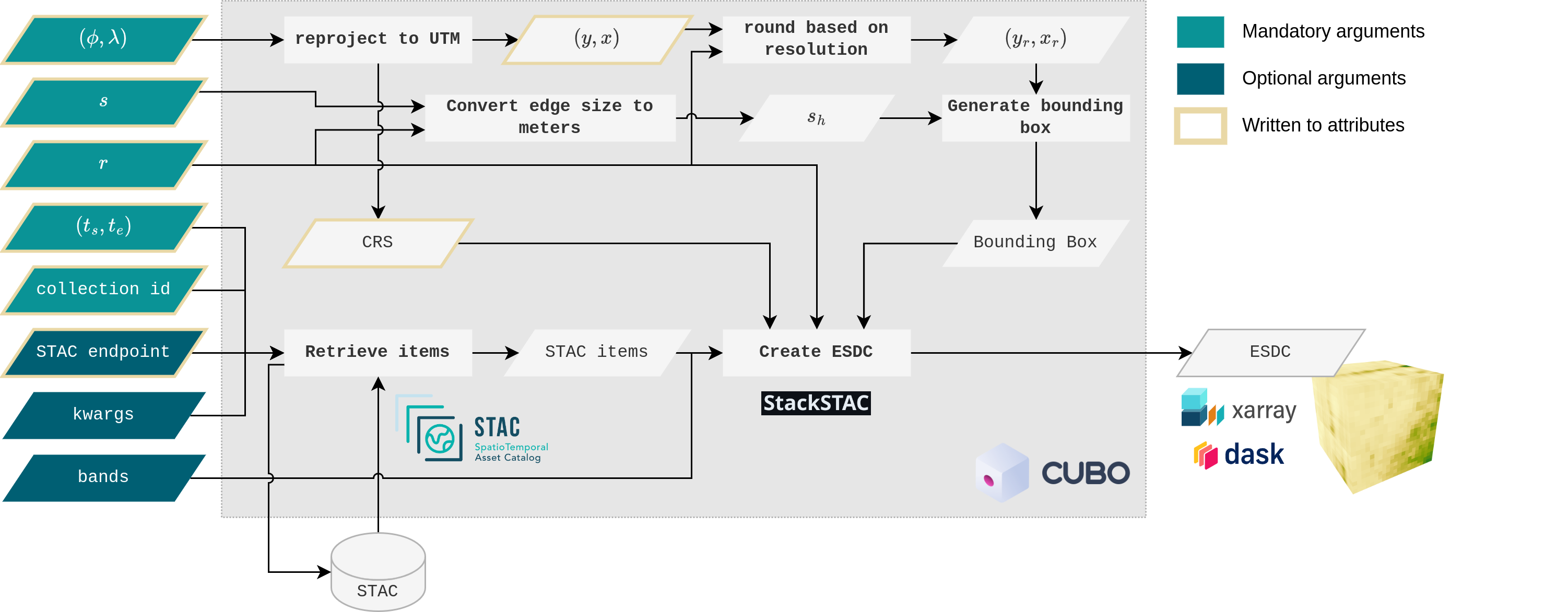}
    \caption{\textbf{Overview of the workflow for building an ESDC inside {\tt cubo}}. This diagram presents a high-level summary of the ESDC construction process.}
    \label{fig:construction}
\end{figure*}

{\tt cubo} builds ESDCs through a series of steps using the above-mentioned parameters as user inputs (Fig.~\ref{fig:construction}):

\subsubsection{Bounding box calculation}
\label{sec:bbox}

{\tt cubo} reprojects $\phi$ and $\lambda$ into their respective Universal Transverse Mercator (UTM) zone coordinates, denoted as $y$ and $x$. Additionally, {\tt cubo} saves the Coordinate Reference System (CRS) as an attribute. These coordinates are then aligned to the nearest pair divisible by the spatial resolution $r$, calculated as:

\begin{equation}
    i_r=r\left\lfloor\frac{i}{r}+0.5\right\rfloor
\end{equation}

Here, $i$ represents either $x$ or $y$, and $i_r$ is the adjusted coordinate. Next, the half-edge size in meters ($s_h$) is calculated as:

\begin{equation}
    s_h=\left\lfloor\frac{sr}{2}+0.5\right\rfloor
\end{equation}

This rounding process ensures that the edge size is an even number. Finally, the bounding box coordinates are determined using:

\begin{equation}
    \begin{bmatrix}
    x_{\textrm{min}} & x_{\textrm{max}}\\
    y_{\textrm{min}} & y_{\textrm{max}}
    \end{bmatrix} = 
    \begin{bmatrix}
    x_r & 1\\
    y_r & 1
    \end{bmatrix}
    \begin{bmatrix}
    1 & 1\\
    -s_h & s_h
    \end{bmatrix}
\end{equation}

Here, the tuples $(x_{\textrm{min}},y_{\textrm{max}})$ and $(x_{\textrm{max}},y_{\textrm{min}})$ denote the upper left and lower right coordinates of the bounding box, respectively.

\subsubsection{ESDC creation}
\label{sec:esdc_retrieval}

In its initial step, {\tt cubo} accesses an STAC catalogue through an endpoint specified by the user, defaulting to the Planetary Computer STAC catalogue's endpoint. Next, {\tt cubo} feeds the bounding box parameters alongside $t_s$ and $t_e$ to {\tt pystac-client} to retrieve all STAC items from a user-defined collection that intersect with these spatio-temporal constraints via a ``search'' operation. ``kwargs'' arguments are an option, enabling users to specify additional query parameters such as cloud cover percentage. Subsequently, these items are transferred to {\tt stackstac}, combined with $r$, the bounding box values, and the CRS. Additionally, the user can define the bands to retrieve. This process culminates in the generation of the ESDC as a ``lazy'' {\tt xarray} object \citep{hoyer2017xarray}, chunked via {\tt dask} \citep{rocklin2015dask}, with the CRS specifically tailored to match the UTM zone associated with $\phi$ and $\lambda$.

\begin{table}[!h]
\footnotesize
\centering
\caption{Global attributes in the ESDC generated by {\tt cubo}.}
    \begin{tabular}{ l p{0.6\linewidth} }
    \toprule
    \textbf{Attribute} & \textbf{Description} \\
    \toprule
    \textbf{collection} & Identifier of the collection within the STAC catalogue. \\ \midrule
    \textbf{stac} & Endpoint of the STAC catalogue used. \\ \midrule
    \textbf{epsg} & EPSG code of the ESDC's CRS, corresponding to a specific UTM zone. \\ \midrule
    \textbf{resolution} & Spatial resolution, denoted by the value of $r$. \\ \midrule
    \textbf{edge\_size} & Edge size of the cube, given by the value of $s$. \\ \midrule
    \textbf{central\_lat} & Central latitude, indicated by $\phi$. \\ \midrule
    \textbf{central\_lon} & Central longitude, represented by $\lambda$. \\ \midrule
    \textbf{central\_y} & UTM coordinate y, corresponding to $y$. \\ \midrule
    \textbf{central\_x} & UTM coordinate x, corresponding to $x$. \\ \midrule
    \textbf{time\_coverage\_start} & Start timestamp of the ESDC, indicated by $t_s$. \\ \midrule
    \textbf{time\_coverage\_end} & End timestamp of the ESDC, given by $t_e$. \\ \bottomrule
    \end{tabular}
\label{table:attributes}
\end{table}

\subsubsection{Attributes writing}
\label{sec:attributes}

After the creation of the ESDC, {\tt cubo} inscribes a set of global attributes on it, as outlined in Table~\ref{table:attributes}. Additionally, {\tt cubo} calculates the Euclidean distance between the coordinates of each pixel in the ESDC and the projected coordinate pair $(y,x)$. This distance array is stored within the ESDC, adhering to the same spatial grids and dimensions, and can be accessed under the coordinate \texttt{cubo:distance\_from\_center}.

\section{Showcase}
\label{sec:showcase}

We illustrate {\tt cubo}'s efficacy through two distinct examples: 1) creating varied ESDCs with different parameters across multiple global locations, and 2) generating a standardised ESDC using different collections with identical parameters in the same location, all sourced from the Planetary Computer STAC catalogue.

\begin{figure*}[!h]
    \centering
    \includegraphics[width=1.0\textwidth]{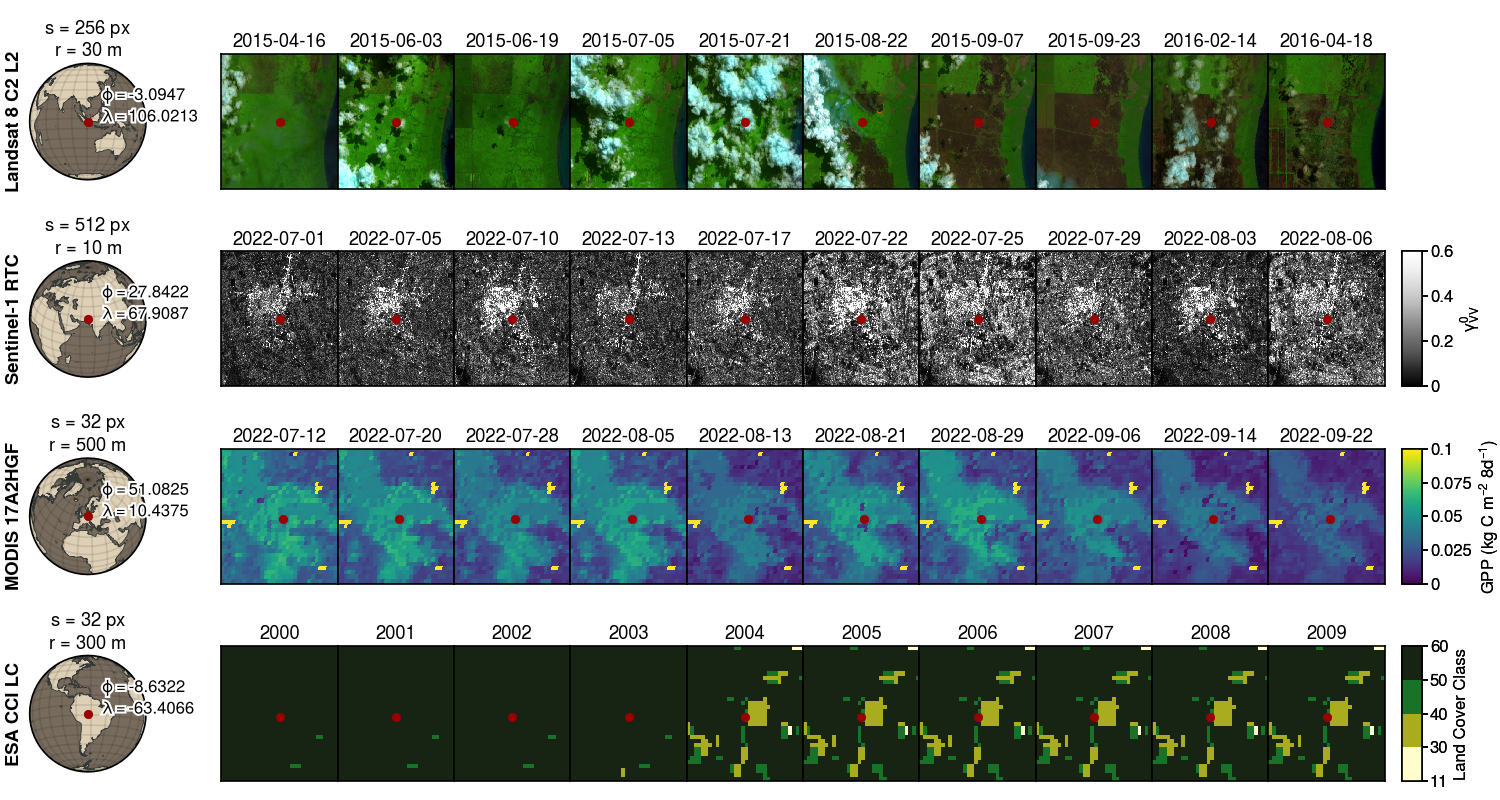}
    \caption{\textbf{Examples of ESDCs generated using {\tt cubo}}. Each row in the figure depicts ESDCs' example timesteps. The first column shows the parameters used and the location of the central coordinates of the ESDC. The first row displays an ESDC generated from Landsat-8 Collection 2 Level 2 data, capturing a fire event in Indonesia. The second row shows an ESDC constructed from Sentinel-1 Radiometrically Terrain Corrected (RTC) data, highlighting a flood event in Pakistan. The third row presents an ESDC based on MODIS 17A2HGF gap-filled 8-day GPP data, depicting a forest area in Germany. The final row features an ESDC created using data from the ESA CCI LC product, focused on Brasil.}
    \label{fig:showcase}
\end{figure*}

In the first scenario, we harnessed various collections specialised for Earth system research (Fig.~\ref{fig:showcase}). The examples were generated across various locations globally, showcasing {\tt cubo}'s versatility in handling data from any region. They also emphasise the importance of spatio-temporal context in a range of applications. For instance, the first three rows in Fig.~\ref{fig:showcase} highlight potential uses in studying climate extremes and their effects on both the natural environment and human society. The first row features a Landsat-8 ESDC with a 30 m resolution, useful for detecting active fires and estimating burned areas. The second row shows a Sentinel-1 ESDC at 10 m resolution, ideal for flood detection and damage assessment. The third row introduces a MODIS-derived Gross Primary Production (GPP) dataset (17A2HGF), with a 500 m resolution, for analysing climate impacts on forest carbon sequestration. Lastly, the fourth row illustrates an annual ESDC from the ESA Climate Change Initiative (CCI) Land Cover (LC) product at 300 m resolution, beneficial as an additional input in various Earth system projects.

\begin{figure}[t]
    \centering
    \includegraphics[width=0.6\textwidth]{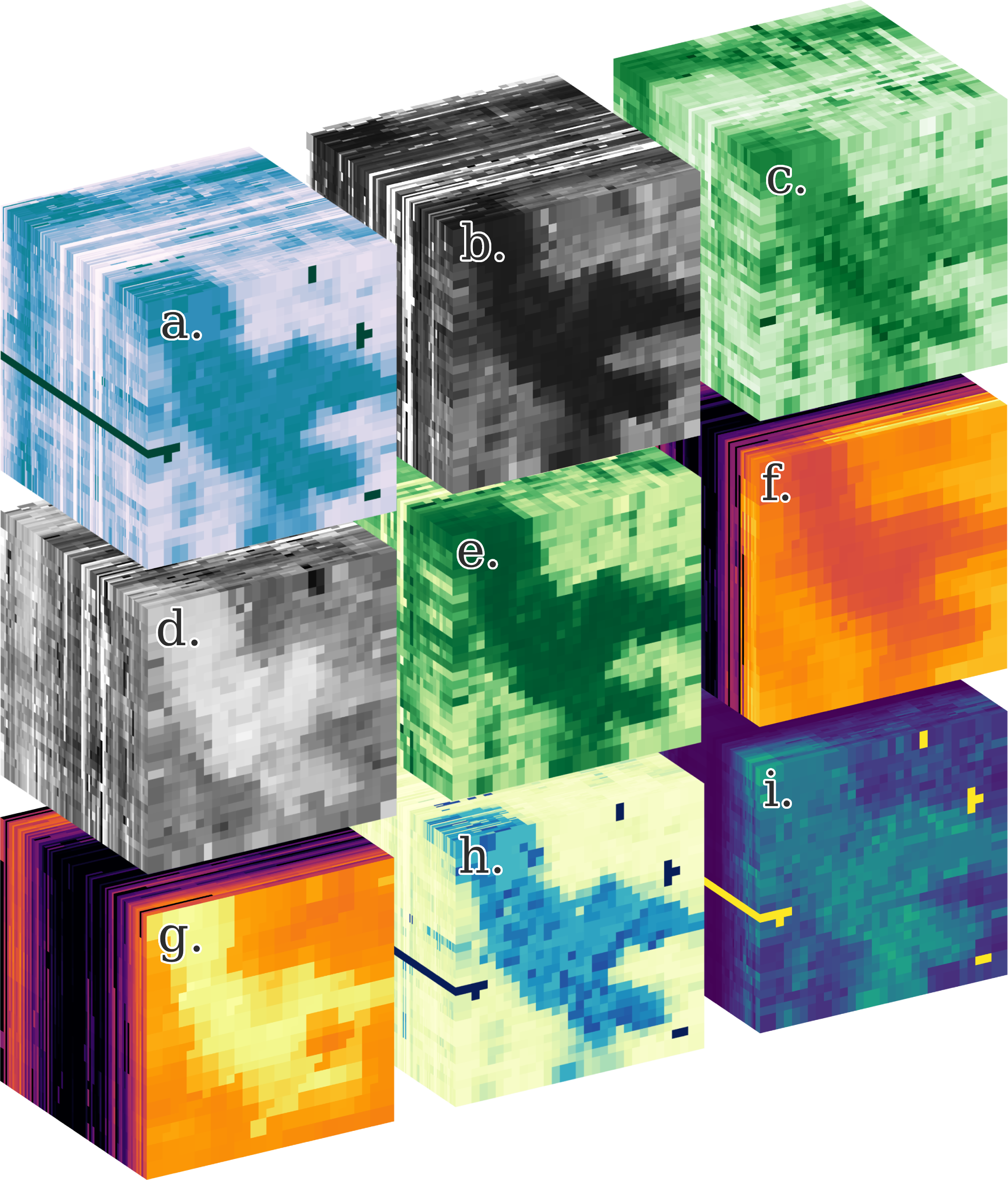}
    \caption{\textbf{Example of ESDCs aligned with a common spatio-temporal grid}. The displayed MODIS datasets are a) FPAR (15A2H), b) Red SR (09A1), c) Enhanced Vegetation Index (EVI, 13Q1), d) NIR SR (09A1), e) Normalised Difference Vegetation Index (NDVI, 13Q1), f) Nighttime LST (11A2), g) Daytime LST (11A2), h) LAI (15A2H), and i) GPP (17A2HGF). The ESDCs were rendered using {\tt lexcube} \citep{soechting2023lexcube}.}
    \label{fig:showcase_single}
\end{figure}

The ultimate goal of ESDCs is to integrate multiple datasets into a singular comprehensive analysis, enhancing our understanding and insights into the Earth system. In the second example, we focused on the same location as the third row of Fig.~\ref{fig:showcase} (DE-Hai Eddy Covariance ICOS site at Hainich National Park, Germany) to create an extensive ESDC from various datasets, all retrieved with a 500 m spatial resolution and a 32-pixel edge size, covering data from 2022-08-01 to 2023-08-01. We gathered MODIS data on Surface Reflectance (SR), Land Surface Temperature (LST), GPP, Leaf Area Index (LAI), Fraction of Photosynthetically Active Radiation (FPAR), and Vegetation Indices (VIs). Fig.~\ref{fig:showcase_single} displays these ESDCs, aligned with the same spatio-temporal grid. It's noteworthy that datasets not matching the requested resolution (e.g. LST, VIs, and SR products) were automatically resampled by {\tt cubo}, defaulting to the nearest neighbours method.

\section{Conclusions}
\label{sec:conclusions}

In this paper, we presented {\tt cubo}, an open-source Python-based tool designed for the straightforward generation of ESDCs on demand. {\tt cubo} simplifies the characterisation of AI-focused ESDCs, requiring minimal user input to create these ESDCs. {\tt cubo} is versatile, and compatible with any COG collection within STAC. We anticipate {\tt cubo} will be instrumental in various analytical processes requiring spatio-temporal context in Earth system research, with a particular emphasis on developing datasets for advanced AI tasks.

\bibliographystyle{abbrvnat}
\bibliography{Bib}

\end{document}